# Fluidic shaping of optical components

Valeri Frumkin[1,2] and Moran Bercovici[2]

[1]Department of Mathematics, Massachusetts Institute of Technology, 77 Massachusetts Ave, Cambridge, MA 02139, USA.

[2]Faculty of Mechanical Engineering, Technion – Israel Institute of Technology, Haifa, 3200003, Israel.

## Abstract

Current methods for fabricating lenses rely on mechanical processing of the lens or mold, such as grinding, machining, and polishing. The complexity of these fabrication processes and the required specialized equipment prohibit rapid prototyping of optical components. This work presents a simple method, based on free-energy minimization of liquid volumes, which allows to quickly shape curable liquids into a wide range of spherical and aspherical optical components, without the need for any mechanical processing. After the desired shape is obtained, the liquid can be cured to produce a solid object with nanometric surface quality. We provide a theoretical model that accurately predicts the shape of the optical components, and demonstrate rapid fabrication of all types of spherical lenses (convex, concave, meniscus), cylindrical lenses, bifocal lenses, toroidal lenses, doublet lenses and aspheric lenses. The method is inexpensive and can be implemented using a variety of curable liquids with different optical and mechanical properties. In addition, the method is scale-invariant and can be used to produce even very large optical components, without a significant increase in fabrication time. We believe that the ability to easily and rapidly create high quality optics, without the need for complex and expensive infrastructure, will provide researchers with new affordable tools for fabricating and testing optical designs.

## Introduction

Lenses are a fundamental component of any optical system, from microscopes to telescopes, holograms, eye-glasses, data storage devices, lasers, and many more. Fabrication of lenses or lens molds relies on mechanical processing such as grinding and machining, followed by polishing of the optical surfaces (1). The demand for high quality surfaces requires specialized and expensive equipment, and the fabrication of non-standard optical surfaces remains challenging (1, 2). It is natural to consider 3D printing technologies as a potential platform for lens prototyping, yet thus far the quality of prints is inadequate for optical applications (3), and complex post-processing is needed in order to achieve required surface quality (4). Furthermore, 3D printing time is proportional to the volume being printed (5) and thus large lenses or a large number of lenses require substantial time to fabricate, making this approach inadequate for industrial scale fabrication.

An alternative approach leverages the smooth liquid-air interface of small polymer droplets to produce high surface-quality lenses (6–8). However, the size of such lenses is restricted by the capillary length of the liquid polymer (representing the relative importance of gravitational to surface tension forces), which for most liquids at standard conditions is < 2.5 mm (9). As the diameter of the droplet approaches the capillary length, gravitational forces become dominant. On top of a horizonal surface, large droplets will be flattened by gravity, resulting in loss of their spherical shape. On an inverted surface, gravity deforms the suspended droplets into an aspherical shape, yet this approach suffers from the same size limitation. (6)

We here present a simple method for rapid fabrication of a variety of high surface-quality lenses that is not limited by size and does not require specialized equipment. The method is based on the injection of a curable lens liquid into an immiscible immersion liquid environment. Under such conditions, in addition to gravity, the liquid polymer experiences a buoyancy force. When the density of the immersion liquid is set to match that of the polymer, neutral buoyancy conditions are achieved, effectively eliminating the dependence on the capillary length. The shape of the resulting lens is determined solely by the injected volume and the geometry of any bounding surfaces in contact with the lens liquid. After the desired shape is obtained, the lens is cured (e.g. polymerized or cross-linked) using standard methods (e.g. photoactivation, thermal setting), and can be removed from the immersion liquid. Due to the natural smoothness of the interface between the lens liquid and the immersion liquid, the quality of the resulting lens is dictated by the molecular polymerization scale and does not require polishing.

We demonstrate the fabrication of plano-convex, plano-concave, bi-convex, bi-concave, bi-focal, meniscus lenses, in addition to non-standard surfaces such as cylindrical lenses, saddle lenses, and aspheric lenses. We also show simple fabrication of doublet lenses based on the combination of several polymers.

## Results

### Theory

Consider a liquid chamber filled with an immersion liquid of density $\rho_{im}$, containing a ring-shaped bounding surface with radius $R_0$ and vertical thickness $d$, positioned at height $h_0$ from the bottom of the chamber (see Fig. 1). A curable lens liquid of density $\rho_{lens}$ which is injected into the bounding surface, is pinned at the boundaries, forming two interfaces, $h^{(t)}(r)$ and $h^{(b)}(r)$, with the immersion liquid (see Supplemental Video S1 for demonstration of the injection process). For the pinning to occur, the lens liquid must have high adhesion to the inner side of the bounding surface.

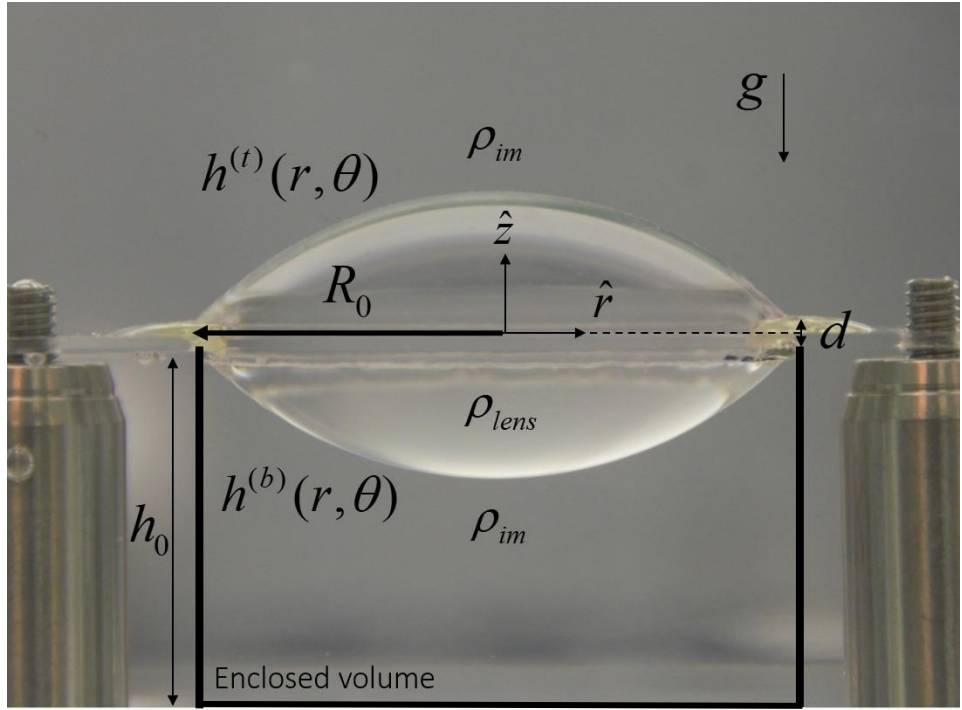

**Fig. 1.** *Schematic illustration of the configuration used for experiments and modeling. The setup consists of a liquid lens injected into a ring-shaped bounding surface which is submerged within an immersion liquid. The outlined segment under the lens liquid represents an enclosed immersion-liquid region whose volume can be controlled for shaping meniscus-type lenses.*

In the above described configuration, the capillary length is defined by $\ell_c = \sqrt{\gamma / \Delta\rho g}$, where $\Delta\rho = |\rho_{lens} - \rho_{im}|$. Thus, if the density of the immersion liquid is set to match that of the lens liquid, the capillary length goes to infinity and interfacial forces dominate over gravity at all scales.

The steady-state shapes of the top and bottom interfaces can be found by minimizing the free energy functional of the system, given by (see Supplementary Information)

$$F = 2\pi\gamma \int_0^{R_0} \left( \sqrt{1+\left(h_r^{(t)}\right)^2} + \sqrt{1+\left(h_r^{(b)}\right)^2} - \frac{\Delta\rho g}{2\gamma}\left(\left(h^{(t)}\right)^2 - \left(h^{(b)}\right)^2\right) + \frac{\lambda_1}{\gamma}\left(h^{(t)} - h^{(b)}\right) + \frac{\lambda_2}{\gamma}h^{(b)} \right) r\, dr \ , \quad (1)$$

where the first two terms under the integral represent surface energy, and the third term represents gravitational energy. The last two terms represent the physical constrains on the system, namely, the volume of the lens liquid and the enclosed volume, with $\lambda_1$ and $\lambda_2$ being the Lagrange multipliers. Minimization of $F$ yields a set of coupled Euler-Lagrange equations for the shapes of the fluidic interfaces, $h^{(t)}$ and $h^{(b)}$,

$$\left(\frac{\Delta\rho g}{\gamma}h^{(t)} - \frac{\lambda_1}{\gamma}\right)r + \frac{rh_{rr}^{(t)} + h_r^{(t)} + \left(h_r^{(t)}\right)^3}{\left(1+\left(h_r^{(t)}\right)^2\right)^{3/2}} = 0, \quad \left(\frac{\Delta\rho g}{\gamma}h^{(b)} - \frac{\lambda_1}{\gamma} + \frac{\lambda_2}{\gamma}\right)r - \frac{rh_{rr}^{(b)} + h_r^{(b)} + \left(h_r^{(b)}\right)^3}{\left(1+\left(h_r^{(b)}\right)^2\right)^{3/2}} = 0, \quad (2)$$

subjected to the boundary conditions

$$h^{(b)}(R_0) = h_0, \ h^{(t)}(R_0) = h_0 + d, \ V_{lens} = 2\pi\int_0^{R_0}(h^{(t)} - h^{(b)})r\, dr, \ V_{enclosed} = 2\pi\int_0^{R_0}h^{(b)}r\, dr. \quad (3)$$

### Spherical Lenses

At neutral buoyancy conditions (i.e., $\Delta\rho = 0$) these equations reduce to the familiar Young-Laplace form, in which case $h^{(t)}(r)$ and $h^{(b)}(r)$ take the shape of perfectly spherical caps.

When the injected volume, $V_{lens}$, is larger than the volume enclosed by the frame, $V_0 = \pi R_0^2 d$, the lens liquid obtains a positive curvature (i.e. a convex lens). Panels (a)-(c) in Fig. 2 demonstrate the ability to control the curvature of a positive lens by varying the injected volume. Panels (g)-(i) present similar results for a taller frame with $V_{lens} < V_0$, in which case the liquid interfaces protrude inward, yielding a negative (concave) lens.

In the case of $|\Delta\rho| > 0$ the up-down symmetry is broken by the additional buoyancy force, resulting in an asymmetric lens. However, as long as $|\Delta\rho| \ll 1$, the lens surfaces can still be well approximated as spheres. Panels (d)-(f) in Fig. 2 show the shapes of a positive, fixed-volume lens, at various densities of the immersion liquid. The up-down symmetry can also be broken by varying the enclosed volume (see Fig.1), thus effectively inflating or deflating the lens liquid, resulting in meniscus-type lenses (see Supplemental Video S2).

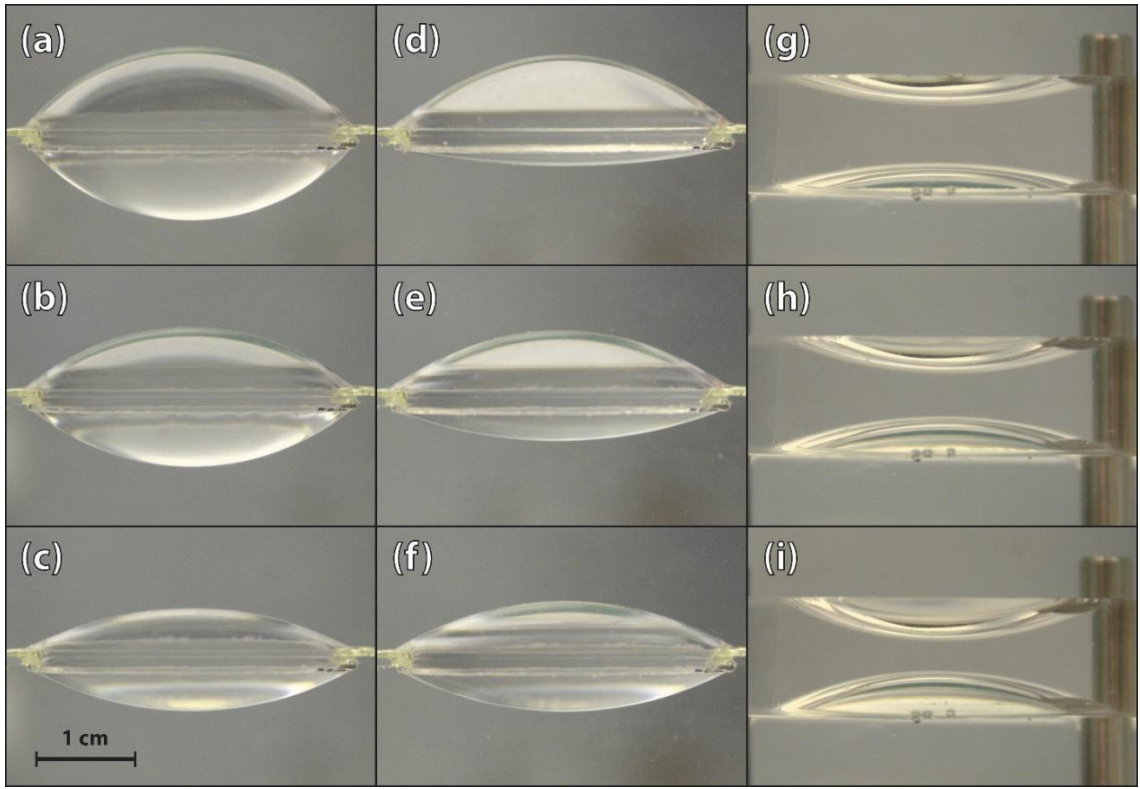

***Fig. 2.*** *Experimental images of spherical lenses produced using ring-shaped bounding surfaces: (a-c) Neutral buoyancy conditions with $V_{lens} > V_0$ result in positive and symmetric spherical lenses, where the lens curvature is dictated by the injected volume. (d-f) Varying slightly the density of the immersion liquid for a fixed volume (here $V_{lens} > V_0$) results in asymmetric spherical lenses. (g-i) Neutral buoyancy conditions with $V_{lens} < V_0$ result in negative and symmetric spherical lenses, where similarly to a-c, the optical the power can be controlled by the injected volume.*

Once the lens liquid assumes its minimum energy shape, it can be solidified by standard methods. In this work, we demonstrate the fabrication of a variety of lenses based on thermal curing of PDMS and on UV curing of an optical adhesive (see Supplementary Information), however, any curable liquid can be used provided that an appropriate immersion liquid can be identified. The lens fabrication is then complete, and not additional processing steps are required.

Fig. 3 presents a collection of solid lenses produced by the fluidic shaping method. Fig. 3a shows the simplest case of a positive spherical lens produced using a ring-shaped bounding surface. Fig. 2b shows a doublet lens produced by a two-step process, where a negative lens was first formed and then used as a base for a positive lens made from a different material (here colored blue for better visualization). Fig. 3c and 3d show respectively a saddle (toroidal) lens and a cylindrical lens created using different volumes injected into a bounding surface composed of a rectangular pad with two perpendicular walls at its sides (see details in Supplementary Information). Fig. 3e shows a bi-focal lens produced by a two-step process, where a lens of one curvature was cut in half and used as part of a new bounding surface for a lens with a different curvature. Since both parts were made using the same polymer (PDMS) they formed a seamless single unit. Fig. 3f shows a negative meniscus lens produced by increasing the enclosed volume below the lens (blue region in Fig. 1). This type of lens is known to reduce spherical aberrations and is standardly used in the eyewear industry. Finally, Fig. 3g shows a 200 mm spherical telescope lens (two orders of magnitude above typical capillary lengths), demonstrating the scale invariance of the fluidic shaping method.

The approach presented here produces lenses with extremely high surface quality. Our atomic force microscopy (AFM) measurements performed across a 20x20 micron area, yield surface roughness values of RMS=1.15 nm and Ra=0.84 nm (see details in Supplementary Information). It is important to emphasize that the surface quality is the direct result of the smoothness of fluidic interfaces and is therefore independent of the lens' shape.

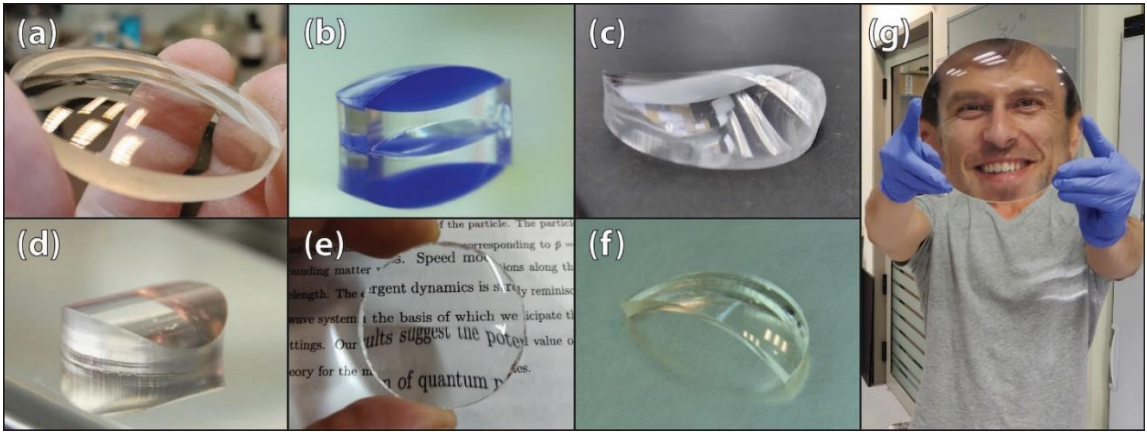

**Fig. 3.** *Images of solid lenses produced using the fluidic shaping method. (a) A positive spherical lens. (b) A doublet lens produced by a two-step process, where a negative lens was used as a bounding frame for a positive lens made from a different material (here colored blue for better visualization). (c) A saddle (toroidal) lens and (d) a cylindrical lens, produced using different lens liquid volumes injected into the same rectangular bounding surface. (e) A bi-focal lens produced by a two-step process, where a first lens was cut in half and used as part of a new bounding surface for a second lens with different curvature. (f) A negative meniscus lens produced by increasing the volume enclosed below the lens. (g) A 200 mm diameter spherical telescope lens.*

### *Aspheric Lenses*

For small deviations from neutral buoyancy, symmetry is broken while the lens surfaces maintain their spherical shape. With further increase in $\Delta\rho$, the liquid interfaces lose their spherical shape and buoyancy effects can no longer be neglected. In this case, we define the following nondimensional variables

$$R = r / R_0, \; H^{(t)}(R) = h^{(t)}(R) / h_c, \; H^{(b)}(R) = h^{(b)}(R) / h_c, \; p_1 = \frac{\lambda_1 h_c}{\gamma \varepsilon^2}, \; p_2 = \frac{(\lambda_1 - \lambda_2) h_c}{\gamma \varepsilon^2}, \quad (4)$$

where $h_c$ is some characteristic vertical deformation length scale (e.g., the maximal value of

$h^{(t)}(r) - h^{(b)}(r)$ ), and $\varepsilon = \left(\dfrac{h_c}{R_0}\right)^2 \ll 1$.

The governing equation now take the form

$$\begin{cases} R\left(BoH^{(t)} - p_1\right)\left(1 + \varepsilon\left(H_R^{(t)}\right)^2\right)^{3/2} + RH_{RR}^{(t)} + H_R^{(t)} + \varepsilon\left(H_R^{(t)}\right)^3 = 0 \\ R\left(BoH^{(b)} - p_2\right)\left(1 + \varepsilon\left(H_R^{(b)}\right)^2\right)^{3/2} - RH_{RR}^{(b)} - H_R^{(b)} - \varepsilon\left(H_R^{(b)}\right)^3 = 0 \end{cases} \quad (5)$$

where $Bo = \dfrac{\Delta\rho g R_0^2}{\gamma}$ is the Bond number.

At leading order in $\varepsilon$, substituting $x = R\sqrt{Bo}$, $P_i = p_i / Bo$, the equations for the top and bottom interfaces take the form

$$xH_{xx}^{(t)} + H_x^{(t)} + xH^{(t)} = P_1 x, \qquad xH_{xx}^{(b)} + H_x^{(b)} - xH^{(b)} = -P_2 x, \qquad (6)$$

where $H^{(i)}$ ($i = t, b$) are the dimensionless functions describing the position of the liquid interfaces, and $x = \dfrac{r}{R_0}\sqrt{Bo}$ is the non-dimensional radial coordinate.

The solutions to these equations are given by

$$H^{(t)} = C_1 \cdot J_0(x) + P_1, \qquad H^{(b)} = C_2 \cdot I_0(x) + P_2, \qquad (7)$$

where $J_0$ and $I_0$ are the zeroth order Bessel and modified Bessel functions, respectively, and $C_1, C_2, P_1, P_2$ are integration constants given by

$$\begin{cases} C_1 = \dfrac{h_c V_{lens} - \pi d Bo + h_c\sqrt{Bo}\left(h_c V_{enclosed} - \pi Bo h_0\right)}{\pi h_c \sqrt{Bo}\left(2 J_1\left(\sqrt{Bo}\right) - \sqrt{Bo} J_0\left(\sqrt{Bo}\right)\right)} \\ P_1 = \dfrac{h_0 + d}{h_c} - C_1 J_0(\sqrt{Bo}) \end{cases}, \begin{cases} C_2 = \dfrac{h_c V_{enclosed} - \pi h_0 Bo}{\pi h_c \sqrt{Bo}\left(2 I_1\left(\sqrt{Bo}\right) - \sqrt{Bo} I_0\left(\sqrt{Bo}\right)\right)} \\ P_2 = \dfrac{h_0}{h_c} - C_2 I_0(\sqrt{Bo}) \end{cases}. \quad (8)$$

Equation (7) allows to obtain a wide range of aspheric Bessel-shaped lenses, by controlling the injected volume, the enclosed volume, and the Bond number of the system, as demonstrated in Supplemental Video S3. It is important to note that these fluidic shapes are stable to significant disturbances even for values of $Bo \sim 1$, as is demonstrated in Supplemental Video S4, removing any time constraints from the polymerization process.

Fig. 4 shows a very good agreement between agreement between the solutions for the top and bottom interfaces given by our theory and experimentally measured ones. In this specific experiment the relevant physical parameters were $\Delta\rho = -6.5\,kg\,/\,mm^3$, $D = 87.2\,mm$, $V_{lens} = 48\,ml$, and $\gamma = 0.02\,N\,/\,m$. We note that our model does not have any fitting parameters, and the results are obtained directly by using actual physical parameters of the system.

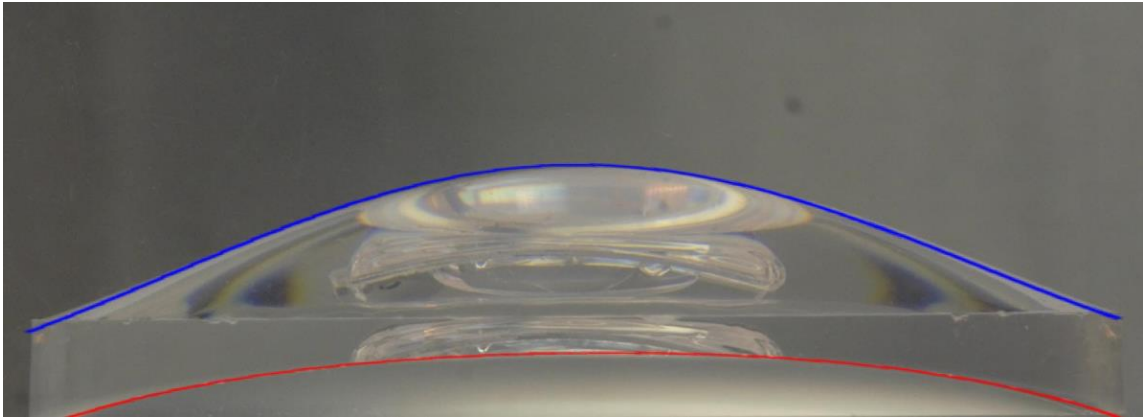

**Fig. 4.** *A comparison of experimental results (image within the liquid container) to our theoretical predictions (overlaid blue and red curves), for $\Delta\rho = -6.5\,kg/m^3$, $D = 87.2\,mm$, $V_{lens} = 48\,ml$, and $\gamma = 0.02\,N/m$, yielding good agreement with no fitting parameters.*

## Conclusion and Outlook

In conclusion, we have demonstrated a method for fluidic shaping of high-quality optical components, allowing for the first-time rapid prototyping of optics. The method is scale-invariant, and unlike its 3D printing counterparts, the required fabrication time is not proportional to the volume being produced, thus allowing to rapidly fabricate components of any size. In addition, the method is compatible with a wide range of curable liquids with various optical and mechanical properties. We identified four degrees of freedom for designing a fluidic optical component – the volume of the component, the enclosed volume, the Bond number, and the shape of the bounding surface. Importantly, the method does not require specialized equipment, and the nanometer-scale surface quality is naturally achieved, without the need for a cleanroom environment, expensive equipment, or complex post-processing (e.g., polishing).

The simplicity and affordability of this method make it a natural candidate to serve as platform for producing affordable eyewear. This is particularly important in countries where the required industrial infrastructure does not exist, preventing 2.5 billion people from access to corrective eyewear, costing the global economy over 1.2 trillion dollars each year (10). Another application of interest is at the opposite scale of size and cost – large high-quality telescope lenses, which are currently produced in laborious and expensive processes. Lastly, one could envision the use of this technology for creation of large fluidic lenses in a microgravity environment (i.e. in orbit) where the immersion liquid becomes redundant. Under these conditions the lens could remain in liquid form, allowing for dynamic control of its curvature.

In this work we focused primarily on an axisymmetric bounding surface, which can yield both spherical and aspheric lenses. However, the examples of the cylindrical and saddle lenses suggest that more general boundary conditions may lead to a wide variety of additional optical shapes. This may be of particular interest to the field of freeform optics where fabrication of non-spherical high-quality lenses remains a challenge (11).

## Materials and Methods

*Immersion liquid:* The immersion liquid was prepared by mixing water with glycerol in varying concentrations. This combination allows to reach any density between 0.997 g/mL for water at room temperature, to 1.263 g/mL for pure glycerol. The precise density can be measured directly by weighing a known volume of the immersion liquid. The simplest way to verify neutral buoyancy conditions is by injecting a small volume of the lens liquid directly into the immersion liquid, without any bounding frame. If the immersion liquid is indeed at neutral buoyancy, the injected lens liquid will take the shape of a spherical drop and will remain stationary, without floating to the surface or sinking to the bottom.

*Lens liquids and curing conditions:* In principle, any curable liquid can be used to form a lens, provided that an appropriate immersion liquid can be identified. In this work, we demonstrate curing of PDMS (Sylgard 184, Dow, MI), and UV adhesive (NOA61, NOA63, NOA81, Norland, NJ) lenses. The PDMS lenses are cured by incubating them at 80 C for 1.5 hr, at 60 C for 4 hr, or at room temperature for 24 hr. The UV adhesive is cured by exposure to light at 365 nm (a 36 W consumer grade nail lamp) for 2-5 min, depending on the thickness of the lens and the specific adhesive chosen. Since both PDMS and Norland adhesives are immiscible in water and have densities between ~1.03 (PDMS) g/mL and ~1.12 g/mL (Norland), the water\glycerol-based immersion liquid allowed us to precisely control the density difference.

## Acknowledgments

We thank Omer Luria and Baruch Rofman for their assistance with AFM measurements and Mor Elgarisi for his help in fabricating the 200 mm lens as well as producing the simulation videos.

**Funding:** This project has received funding from the European Research Council under the European Union's Horizon 2020 Research and Innovation Programme, Grant agreement 678734 (MetamorphChip);

# Fluidic Shaping of Optical Components

**Varying the shape of the bounding surface**

Figure S1 presents a schematic illustration of several configuration of bounding surfaces used in this work. The central configuration considered here is that of a ring-shaped bounding surface (Fig S1a,b). For $\Delta\rho = 0$ this shape allows to obtain convex, concave, and meniscus type spherical lenses, while for $\Delta\rho \neq 0$ Bessel-shaped aspheric lenses can be obtained.

The fluidic shaping method can also be used to produce non-axisymmetric optical components. For example, a bounding surface in the shape of a rectangular pad with two sidewalls can be used to produce a cylindrical lens (Fig. S1c). The injected lens-liquid wets the pad and the sidewalls, and the boundary conditions are set by the pinning of the contact line at the edges of the pad. The cylindrical shape is obtained for the corresponding volume of the injected lens-liquid. If the injected volume is less than that required by for a cylindrical lens, the same boundary conditions result in a saddle (toroidal) lens. This is of course only a subset of the possible components achievable by this method and combining more general boundary conditions with variation of the Bond number, can produce a very wide range of optical shapes.

**Surface roughness**

Surface profile measurements were obtained using reflective digital holographic microscopy, (DHM-R, LynceeTec) with a 10X objective, of a spherical lens produced from NOA 81 UV adhesive. The measured profile shows excellent agreement with a least square fit of a circular section, with a maximum deviation of 15 nm over an area of 500x500 microns (<u>Fig.</u> S2). To obtain precise measurements of surface roughness we sent several samples for measurement by atomic force microscopy, and measured each at several locations. The measurements indicated roughness values of RMS=1.15 nm and Ra=0.84 nm for the best sample, and RMS=3.16 nm and Ra=1.89 nm for the worst ones. All measurements were performed over a 20x20 micron area.

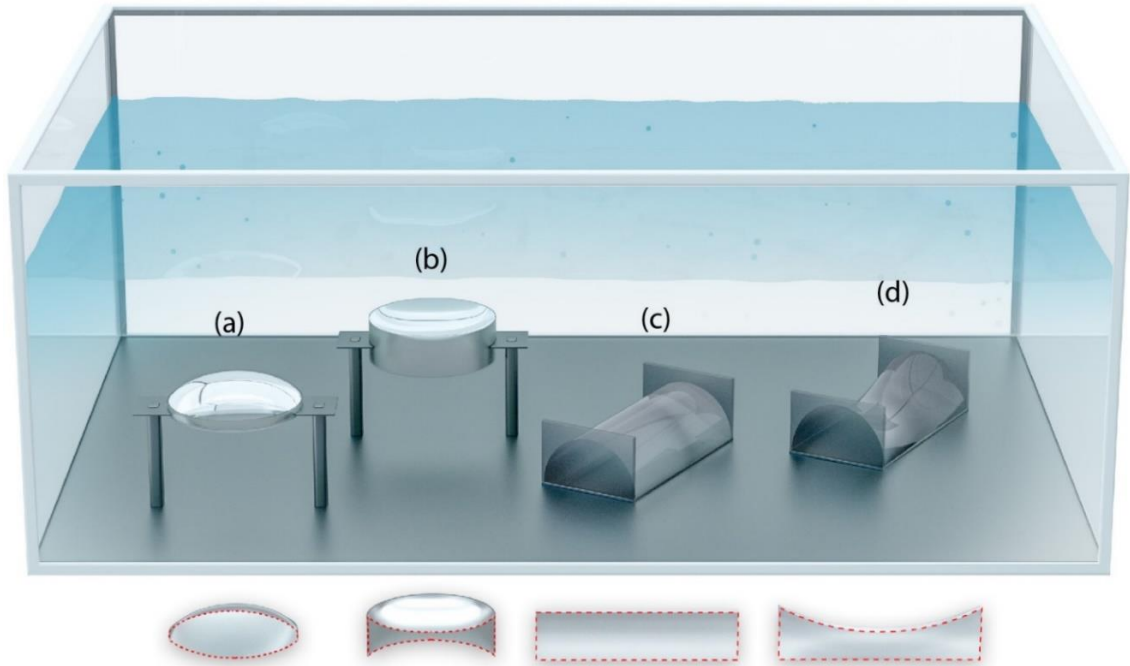

**Fig. S1.** *Schematic illustration of the experimental setup for the fluidic shaping method, presenting several bounding surfaces considered in this work. (a) A thin ring-shaped surface used to produce a positive spherical lens. (b) A thick ring-shaped surface used to produce a negative spherical lens. (c) A rectangular pad with two walls, used to produce a cylindrical lens. (d) Same as shape as in (c) with less injected volume, results in a saddle lens.*

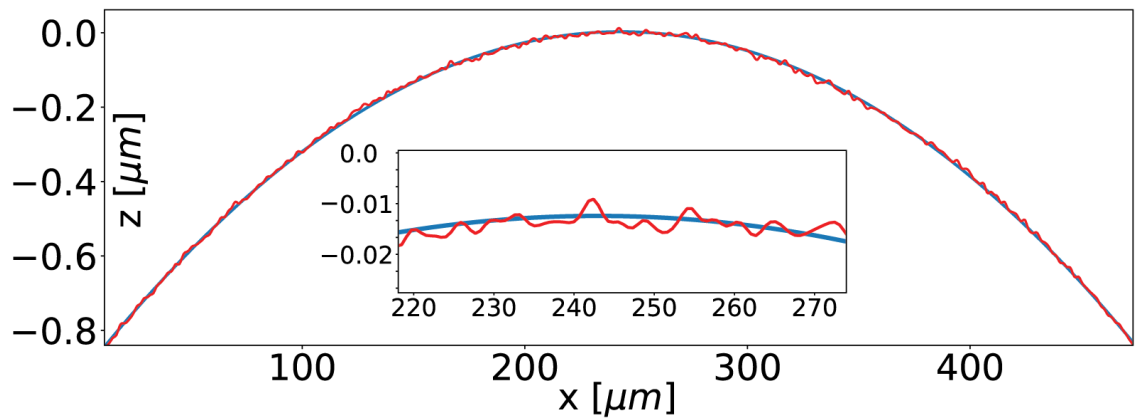

**Fig. S2.** *A surface profile measurement of a spherical lens, obtained by reflective digital holographic microscopy (red), compared to a least square fit of a circular section (blue). The maximum deviation is 15 nm.*